\documentclass[%
 reprint,
nofootinbib,
 amsmath,amssymb,
 aps,
]{revtex4-2}

\usepackage{graphicx}% Include figure files
\usepackage{dcolumn}% Align table columns on decimal point
\usepackage{bm}% bold math
\usepackage{tikz}
\usetikzlibrary{snakes}
\usetikzlibrary{calc,decorations.markings}
\usepackage[compat=1.1.0]{tikz-feynman}
\usepackage{subcaption}
\usepackage{feynmf}
\usepackage{hyperref}
\usepackage{cleveref}
\usepackage{float}
\usepackage{physics}
\usepackage{pgfplots}
\usepackage{lipsum}
\usepackage{ulem}
\usepackage[colorinlistoftodos]{todonotes}
\Crefname{equation}{Eq.}{Eqs.}
\Crefname{figure}{Fig.}{Figs.}
\Crefname{definition}{Def.}{Defs.}

\newcommand{\BigO}[1]{\mathcal{O}( #1 )}

\begin{document}

\preprint{APS/123-QED}

\title{Accelerated Particle Detectors with Modified Dispersion Relations.}% Force line breaks with \\

\author{Paul C.W. Davies}
\email{paul.davies@asu.edu}
\author{Philip Tee}
\altaffiliation[Also at ]{p.tee@sussex.ac.uk, \\Department of Informatics, \\University of Sussex, Falmer, UK.}
\email{ptee2@asu.edu}
\affiliation{The Beyond Center for Fundamental Science,  Arizona State University, Tempe AZ 85287, USA}

\date{\today}% It is always \today, today,

\begin{abstract}
There is increasing interest in discrete or ‘pixelated’ spacetime models as a foundation for a satisfactory theory of quantum gravity. 
If spacetime possesses a cellular structure, there should be observable consequences: for example, the vacuum becomes a dispersive medium. 
Of obvious interest are the implications for the thermodynamic properties of quantum black holes. 
As a first step to investigating that topic, we present here a calculation of the response of a uniformly accelerating particle detector in the (modified) quantum vacuum of a background pixelated spacetime, which is well known to mimic some features of the Hawking effect. 
To investigate the detector response we use the standard DeWitt’s treatment, with a two-point function modified to incorporate the dispersion. 
We use dispersion relations taken from the so-called doubly special relativity (DSR) and Ho{\u r}ava-Lifshitz gravity.
We find that the correction terms retain the Planckian nature of particle detection, but only for propagation faster than the speed of light, a possibility that arises in this treatment because the dispersion relations violate Lorentz invariance. 
A fully Lorentz-invariant theory requires additional features; however, we believe the thermal response will be preserved in the more elaborate treatment.
\end{abstract}

\maketitle

%\tableofcontents

\section{Introduction and Background}
\label{sec:introduction}
Much of theoretical physics is formulated on the assumption that spacetime is continuous and maps to the real numbers. While this is an obvious idealization, convenient for calculation, it rarely presents difficulties. 
However, in quantum field theory, spacetime continuity leads to divergences that must be evaded by renormalization – an ad hoc procedure. 
Worse still, many formulations of quantum gravity are non-renormalizable.

What happens if spacetime continuity is discarded? 
There is a rich history of models in which spacetime emerges in the macroscopic limit from some sort of discrete or pixelated substructure \cite{ambj2002lorentzian,ambj1997quantum,trugenberger2015quantum,trugenberger2017combinatorial,tee2020dynamics,tee2021canonical,rovelli2014covariant}. 
Quantum gravity defines a fundamental length scale, the Planck length $l_p  \equiv \sqrt{\frac{\hbar G}{c^3}}=1.616 \times 10^{-35}~m$, which provides a natural measure of the ‘pixel’ size, and a natural cut-off energy for any incipient ultra-violet divergence.
Although ‘pixelating’ spacetime, i.e. replacing real numbers by a countable infinity, is quite probably a placeholder for some more nuanced micro-structure, or pre-geometry \cite{wheeler2018information}, it is interesting to explore the impact of this simple modification to determine whether there are any observable consequences. 
One of the most basic spacetime structures, and the most intensively studied, is the black hole. 
How might spacetime discreteness affect its properties, such as Hawking radiation? 
As a first step to addressing this question, we here investigate the response of an accelerating particle detector, the so-called Davies-Fulling-Unruh effect, known to mimic Hawking radiation, in a pixelated spacetime background. 

Although the existence of a fixed fundamental length scale violates Lorentz invariance, Minkowski space may be retained if momentum space acquires non-zero curvature, a construction known as Doubly Special Relativity (DSR) \cite{amelino2001testable,amelino2002relativity,ambj2002lorentzian}. 
DSR may be generalized to curved spacetime by replacing locally Minkowski space with locally de Sitter space \cite{aldrovandi2007sitter}. 

The effect of introducing curvature in momentum space is to render the vacuum a dispersive medium in which different frequency light waves propagate at different speeds. 
This can be modelled by adding successively higher powers of the three-momentum $p$ to the standard energy-momentum dispersion relations, 
\begin{equation}\label{eqn:massive_disp}
    E^2 = p^2 + m^2 + \frac{\kappa_1}{M_p} p^3 + \frac{\kappa_2}{M^2_p} p^4 + \dots \text{,}
\end{equation}
where the dimensionless kappa coefficients are arbitrary at this stage.
The dispersion relation \Cref{eqn:massive_disp} on its own puts the theory in violation of Lorentz invariance, a defect that may, however, be remedied in a more elaborate treatment that involves  changing the momentum measure  \cite{amelino2014planck,mattingly2005modern}.
We defer the complete DSR treatment in this paper and restrict our analysis to the effect of the dispersion relation alone on the response of a particle detector.

If the vacuum really is dispersive, then there should be observable effects. 
Indeed, it is known that the $p^3$ term produces a correction to the black hole entropy which is usually discarded by setting $\kappa_1$ to zero \cite{amelino2004severe}, a practice to which we will adhere in this paper, focusing instead on the consequences of non-zero $p^4$ and $p^6$ terms (the latter arising in Ho{\u r}ava-Lifshitz theories of gravity (HLG) \cite{hovrava2009quantum}).
Modifications to the dispersion relations in turn alter the Green’s functions of any quantized field theory, which we have previously explored in the context of the effect of these changes on the bending of light by massive bodies \cite{tee2022fundamental}. 
The same Green’s functions can be adapted for use in calculating the response of a DeWitt-Unruh particle detector.
The principle results of this calculation are presented in \Cref{sec:accelerated}, with the details covered in the appendices.

A similar calculation was performed by Rinaldi \cite{rinaldi2008superluminal}, but crucially only the case of positive $\kappa_2$ was considered, and higher order terms in $p$ were not considered.
The approach taken there was to treat the higher order terms in the dispersion relationship perturbatively, whereas our calculation is more direct.
For positive $\kappa$ we obtain a similar result to Rinaldi, but our method is more general and able to accommodate negative as well as positive correction terms.

The layout of the paper starts with an overview of the effect of extra terms on the propagation of waves through the vacuum in \Cref{sec:dispersion}, and establish that $\kappa$ controls whether propagation is sub or superluminal.
In \Cref{sec:greensfunctions} we present the modified Hadamard functions that we use in \Cref{sec:accelerated} to compute the corrections to the detector response functions.
We then briefly consider the effect of higher order terms in the dispersion relations in \Cref{sec:Horava_lifshitz}, before concluding with a discussion of the impact of these results on other semi-classical phenomena in \Cref{sec:conclusion}.

\section{Modified dispersion relation and refractive index}
\label{sec:dispersion}
We restrict the discussion here to the modified dispersion relation,
\begin{equation}\label{eqn:dispersion_massive}
    E^2=p^2+\kappa \eta^2p^4 +m^2\text{,}
\end{equation}
where $\eta$ is a characteristic length scale related to the pixelation scale, and $\kappa$ controls the sign of the correction. 
In the massless limit with units $\hbar=1,E=\omega,$ and $p=k$ we note that a scalar field $\phi(t,x)$ obeys the modified free space wave equation,
\begin{equation}\label{eqn:eqom}
    \pdv[2]{\phi}{t} - \pdv[2]{\phi}{x} + \kappa \eta^2 \pdv[4]{\phi}{x} = 0 \text{,}
\end{equation}
which has a set of solutions $\phi(t,x) = e^{\pm(\omega t - k \vdot x)}, e^{\pm i(\omega t -k \vdot x)}$, shown to be complete by computation of the Wronskian.
The key result is that, although \Cref{eqn:eqom} has wavelike solutions, the vacuum itself is dispersive and possesses a refractive index $n(k)$ given by,
\begin{equation}\label{eqn:refractive}
        n(k)=\frac{\sqrt{1+\kappa \eta^2 k^2}}{1+2\kappa \eta^2 k^2} \text{,}
\end{equation}
a result that extends to spin 1 (photons) and (linearized) spin 2 (gravitons) too, and broadly follows the analysis by Myers {\sl et al} \cite{myers2003ultraviolet}.
The fact that the speed of wave propagation of the waves depends on the frequency introduces some peculiar features. 
Note that for $\kappa > 0$ high frequency radiation propagates faster than low frequency radiation, and all waves propagate at a speed $> 1$, i.e. faster than the speed of light in the unmodified case.  
For the case $\kappa \leq 0$ the propagation speed is $\leq  1$ and slows as $k$ increases, albeit slowly.  
There is a singularity at $k = \frac{1}{\eta \sqrt{2 \kappa }}$, which is determined by the pixalation length, which might naturally be associated with the Planck length. 
However, for the purposes of this paper, we shall sidestep the choice of length scale, as the key issue from the above discussion is the sign rather than the value of $\kappa$.  

There is the obvious problem of reconciling superluminal propagation with relativistic causality. 
This is of course expected since the dispersion relations from DSR and HLG explicitly violate Lorentz invariance.
Locally Lorentz invariant discrete spacetimes theories can be constructed by introducing curvature in momentum space \cite{amelino2010threshold}.
We think our principal result will remain true in the more general case. 

\section{Green's functions in position space}
\label{sec:greensfunctions}
In the massive case, \Cref{eqn:dispersion_massive} leads to a modified propagator in n-dimensional momentum space, and associated Feynman diagram \cite{tee2022fundamental},
\begin{figure}[H]
	\raggedright
        \setlength{\parindent}{1.2cm}
        \begin{fmffile}{prules}
        \begin{fmfgraph*}(55,40)
            \fmfleft{o1}
            \fmfright{o2}
            \fmf{dashes_arrow, label=$p$}{o1,o2}
            \fmflabel{}{o1}
            \fmflabel{\Large $=\frac{1}{p_0^2-p^2(1+\kappa \eta^2 p^2)-m^2 + i\epsilon}$,}{o2}
        \end{fmfgraph*}
        \end{fmffile}
\end{figure}
and the position space propagator, 
\begin{equation}\label{eqn:lorentz_prop}
    G(t,x;t',x')=\int\limits_{-\infty}^\infty \frac{\dd^n p}{(2\pi)^n} \frac{e^{-i[p_0(t-t') -\va{p}\vdot(\va{x}-\va{x'})]}}{p_0^2-p^2(1+\kappa \eta^2 p^2) - m^2 } \text{.}
\end{equation}
The $p_0$ integral is performed using a contour integral noting that the poles occur at $p_0^2=p^2(1+\kappa \eta^2 p^2) + m^2$,  with the choice of contour related to the relevant two-point functions in the standard manner \cite{birrell1984quantum}. 
For our treatment  we require the positive and negative frequency Hadamard functions, 
\begin{equation}\label{eqn:base_prop}
    G(t,x;t',x')=\frac{1}{(2\pi)^{n-1}}\int\limits_{-\infty}^\infty \frac{\dd^{n-1} p}{2E_p} e^{-i[E_p(t-t') -\va{p}\vdot(\va{x}-\va{x'})]} \text{,}
\end{equation}
where $E_p=\pm \sqrt{p^2(1+\kappa \eta^2 p^2) + m^2}$.

Note that we use the standard Lorentz invariant measure from conventional quantum field theory in this integral, and not one modified by attributing a curvature to momentum space, in accordance with the simpler Lorentz violating \sout{version of the} theory we are using. 
The integration in \Cref{eqn:base_prop} is complicated, but we require only the small corrections $\BigO{\kappa\eta^2}$ here. 
The details of these calculations are contained in the Appendices. 
In the massless limit, $3 + 1$ case, we find for time-like separation, 
\begin{equation}\label{eqn:final31timelike}
    D^+(t,r)=\frac{-\theta(\sigma^2)}{4\pi^2(\sigma^2 - i\epsilon)}\left \{ 1-\frac{\kappa \eta^2}{(\sigma^2-i\epsilon)}\right \} \text{,}
\end{equation}
and for the (retarded) space-like case,
\begin{equation}\label{eqn:final31spacelike}
    D^-(t,r)=\frac{\theta(-\sigma^2)}{4\pi^2(-\sigma^2 + i\epsilon)}\left \{ 1+\frac{\kappa \eta^2}{(-\sigma^2+i\epsilon)}\right \} \text{.}
\end{equation}
Here we denote by $\sigma^2=t^2-r^2$ the invariant interval, with $\theta(x)$ being  the normal sign function.
In the case of $\eta=0$ we recover the standard result \cite{hong2010analytic}.

\section{Accelerated Detectors}
\label{sec:accelerated}
Following \S 3.3 of \cite{birrell1984quantum}, we can use the correction to the propagator to investigate the response of an accelerated particle in pixelated spacetime by leveraging the modifications to dispersion relations \sout{ from DSR}.
We consider a particle detector under uniform inverse acceleration $\alpha$, executing a hyperbolic path in Minkowski space parameterized by proper time $\tau$ according to,
\begin{align*}
    x&=y=0, ~z=(t^2+\alpha^2)^{\frac{1}{2}}, ~\alpha=\text{const, such that,}\\
    z&=\alpha \cosh{\frac{\tau}{\alpha}}, ~t=\alpha\sinh{\frac{\tau}{\alpha}} \text{,}
\end{align*}
The standard DeWitt-Unruh detector \cite{unruh1976notes,davies1975scalar,dewitt1979quantum} leads to the transition probability per unit proper time, 
\begin{equation}\label{eqn:transprob}
    c^2 \sum\limits_E \abs{\bra{E}m(0) \ket{E_0}}^2 \int\limits^\infty_{-\infty} \dd (\Delta \tau) ~ e^{i(E-E_0)\Delta \tau}D^+ (\Delta \tau) \text{,}
\end{equation}
and the detector response function,
\begin{equation}\label{eqn:detector_integral}
    \mathcal{F}(E)=\int\limits^\infty_{-\infty} \dd (\Delta \tau) ~ e^{i(E-E_0)\Delta \tau}D^+ (\Delta \tau) \text{.}
\end{equation}
The response function is modified by the correction term in the propagator (see \Cref{sec:appendix_c}) to yield,
\begin{widetext}
\begin{equation}\label{eqn:detector_final}
    \mathcal{F}(E) = \left \{ \left ( 1+\frac{\kappa\eta^2}{6 \alpha^2}\right ) \frac{E-E_0}{e^{2\pi(E-E_0)\alpha} -1 } + \left (\frac{\kappa\eta^2}{6}\right) \frac{(E-E_0)^3}{e^{2\pi(E-E_0)\alpha} -1 } \right \} \text{}
\end{equation}
\end{widetext}
which retains a thermal character. 
We also show (\Cref{sec:appendix_c}) that the modified Green’s function remains periodic in imaginary time with period $2\pi\alpha$, as in the standard (unmodified) treatment. 

Although the modification to the detector response looks innocuous at first sight, it contains some troublesome features. 
If $\kappa < 0$ (subluminal propagation) the transition probability becomes negative for $\abs{\kappa} > \frac{6\alpha^2}{\eta^2}$. 
This may point to a breakdown of unitarity or some other pathology with the theory, but only in cases of extreme acceleration (very small $\alpha$). Our results would still be valid over a wide range of accelerations. Alternatively, one might simply rule out values of $\kappa < 0$, corresponding to subluminal propagation, as unphysical.

In the case of $\kappa > 0$, (superluminal propagation),the detector response function has no worrying negative terms that would produce a pathological transition probability.
In some sense this is consistent with the fact that the calculation is not Lorentz invariant, and so by definition places no restriction of propagation velocities.
This was essentially the case considered by Rinaldi \cite{rinaldi2008superluminal}, and we need to perform the full Lorentz invariant calculation to investigate whether this behavior survives.

\section{Effect of higher terms on the result}
\label{sec:Horava_lifshitz}
We now briefly consider higher order terms in the dispersion relations. 
It was remarked in \Cref{sec:introduction} that (HLG) can introduce a $p^6$ term into the dispersion relations. 
The central idea of HLG gravity is an anisotropic scaling of space and time, achieved by introducing a scaling parameter $b$ and critical exponent $z$, such that $\va{x} \rightarrow b \va{x}$, and $t \rightarrow b^z t$. 
As a quantized theory this becomes power counting renormalizable at a value of $z=3$ \cite{hovrava2009quantum}, and we focus on the modifications this makes to the dispersion relations. 
The anisotropic scaling introduces new diffeomorphisms, that in turn introduce additional terms to the gravitational action, in particular a potential term that alters the dispersion relations. 
At the high energy UV limit these modifications suggest the following form of the dispersion relations,
\begin{equation}\label{eqn:Horava_disp}
    E^2=p^2+m^2 +\kappa \eta^2 p^4 - \kappa \eta^4 p^6 \text{.}
\end{equation}
The additional $p^6$ term is explicitly set to have the opposite sign to the $p^4$ term, as required to ensure that the sign of $\kappa$ dictates whether propagation is super or subluminal \cite{amelino2010threshold,alexandre2015effective}.

We can formulate an associated equation of motion to the dispersion relationship \Cref{eqn:Horava_disp} to determine the refractive index of the vacuum. 
This modifies the equation of motion \Cref{eqn:eqom} to
\begin{equation}\label{eqn:eqom_hl}
    \pdv[2]{\phi}{t} - \pdv[2]{\phi}{x} + \kappa \eta^2 \pdv[4]{\phi}{x} -\kappa\eta^4 \pdv[6]{\phi}{x}= 0 \text{,}
\end{equation}
with the corresponding momentum-dependent refractive index,
\begin{equation}\label{eqn:refractive_hl}
        n(k)=\frac{\sqrt{1+\kappa \eta^2 k^2 + \kappa\eta^4k^4}}{1+2\kappa \eta^2 k^2+3\kappa\eta^4k^4} \text{.}
\end{equation}
As before, for $\kappa > 0$ propagation is superluminal, and $\kappa < 0$ restores subluminal propagation.

The impact of this new dispersion relation on the detector response function is straightforward to calculate as detailed in \Cref{sec:appendix_d} and we obtain the following modification to the detector response function,
\begin{widetext}
\begin{equation}\label{eqn:detector_Horava}
    \mathcal{F}(E) = \left \{ \left ( 1+\frac{\kappa\eta^2}{6 \alpha^2} -\frac{3\kappa'\eta^4}{5\alpha^4}\right ) \frac{E-E_0}{e^{2\pi(E-E_0)\alpha} -1 } + \left (\frac{\eta^2[\kappa+12\kappa'\eta^2]}{6}\right) \frac{(E-E_0)^3}{e^{2\pi(E-E_0)\alpha} -1 } + \left (\frac{\kappa'\eta^4}{10}\right) \frac{(E-E_0)^5}{e^{2\pi(E-E_0)\alpha} -1 } \right \} \text{.}
\end{equation}
\end{widetext}
The calculation introduces a new sign term $\kappa'=\kappa\left ( \frac{3}{4}\kappa + 1 \right )$, but for $\kappa= \pm 1$ it has the same polarity as the original term (i.e. for $\kappa=\pm 1, \kappa'=\pm1$).

The first thing to note is that the additional contributions arising from the $p^6$ term retain the Planck factor, and so preserve the thermal nature of the response.
However, as in \Cref{eqn:detector_final}, we have problematic features that arise in the case of subluminal propagation where $\kappa,\kappa'<0$.
The new terms in the detector function pick up a coefficient of $\eta^4$ from the dispersion relation, and are therefore much smaller in magnitude than the $p^4$ terms.
The only hope for restoring positive definiteness would arise from the $\frac{3\kappa'\eta^4}{5\alpha^4}$ coefficient of the first term overwhelming the $\frac{\kappa\eta^2}{6 \alpha^2}$ from the original calculation.
If we assume $\eta \sim \frac{1}{M_p}$, this would only be the case for very small accelerations in the order of a few $ms^{-2}$ or $\alpha \sim c$.
At this point the absence of a factor of $\alpha$ in the second Planck term would cause that to dominate the sign of $\mathcal{F}(E)$ and give it an overall negative value.
The converse argument for superluminal propagation ($\kappa,\kappa' > 1$) confirms that the detector response function remains positive across the full range of $\alpha$ in this case.
 
We can therefore conclude that the detector response function remains potentially problematic for subluminal propagation, and the possibility exists that the transition probability will then also be negative. 
On dimensional grounds, for higher powers of $p$ in the dispersion relation, the corrections will acquire increasing powers of $\eta$ and follow the same pattern for their contribution to the response function.
One would expect that similar arguments would apply to those corrections, and the detector response function will cease to  be positive definite in the case of subluminal propagation. 
We conclude that any route to cancellation of the $p^4$ term from higher order terms in dispersion relations appears shut off.

\section{Conclusion and discussion}
\label{sec:conclusion}
The principal result of this paper is the discovery of severe pathologies  in the calculation of the Davies-Fulling-Unruh effect in the case of discrete spacetime theories, unless we are willing to accept superluminal propagation.
Although we use the modified dispersion relations derived from DSR, a fully Lorentz-invariant treatment that accommodates a fundamental length, requires additional features in the theory that we shall address in a future paper.
The close relationship between the Davies-Fulling-Unruh effect, systems of accelerating mirrors and Hawking radiation makes this result even more interesting. 
Adding higher order terms in the dispersion relations does not provide a remedy to the issue. 
In fact due to the relative size of contribution at higher powers of $p$ and the appearance of powers of inverse acceleration, the likelihood of cancellation from higher order momentum terms in the dispersion relationship appears low. 
One is led to the conclusion that the coarse-grained structure of spacetime may significantly modify, or even suppress, thermal vacuum effects for certain parameter ranges. 
It is well recognized that introducing a simple cut-off in trans-Planckian modes occasioned by the existence of a fundamental length is problematic for some derivations of the Hawking effect \cite{hawking1975particle,jacobson1991black}. Further investigation of how vacuum dispersion affects the calculation of the Bogoliubov transformation in the black hole radiance calculation may clarify these issues.

In particular our principal conclusion concerning pathologies in the detector response function arise when considering  subluminal propagation may not survive the re-imposition of Lorentz invariance.
The question of whether this is indeed the case is the subject of ongoing investigation.

\appendix

\section{Computation of position space Green's function in 3+1 spacetime}
\label{sec:appendix_b}
\subsection{First method of computation}
\label{sec:app_b1}
We start with the massless propagator as defined in \Cref{eqn:base_prop}, which we convert to $n$-dimensional spherical polar momentum space to obtain,
\begin{equation*}
    D(t,r)=\frac{2(\pi^{\frac{n-2}{2}})}{(2\pi)^{n-1}\Gamma(\frac{n-2}{2})}\int\limits_0^\pi \sin^{n-3} \theta ~\dd \theta \int\limits_0^\infty \frac{\dd^n p}{2E_p} e^{-i(E_p t -p r)} \text{.} 
\end{equation*}
The angular integral can be performed by analytic continuation of the standard result from \cite{gradshteyn2014table} (pg. 346. 3.387 (1) ),
\begin{equation*}
    \int\limits_0^\pi \sin^k \theta e^{i p r \cos \theta} ~\dd \theta = \sqrt{\pi}\left( \frac{2}{pr}\right)^{\frac{k}{2}} \Gamma \left( \frac{k+1}{2}\right ) J_{\frac{k}{2}}(pr) \text{,}
\end{equation*}
where $J_n(x)$ are the Bessel functions of order $n$.
To remove clutter in what follows, we define,
\begin{equation}\label{eqn:zeta_def}
    \zeta=\frac{1}{2(2\pi)^{(n-1)/2}r^{(n-3)/2}} \text{.}
\end{equation}
Using this result and inserting our modified form for $E_p$, we have the following integral to perform,
\begin{equation}\label{eqn:base_integral}
    D(t, r) = \zeta \int\limits_0^\infty \dd p ~\frac{p^{\frac{n-3}{2}}}{(1+\kappa \eta^2p^2)^{1/2}}J_{\frac{n-3}{2}}(p r)e^{-ip(1+\kappa \eta^2p^2)^{1/2} t}\text{.}
\end{equation}
Exploiting the fact that $\kappa\eta^2$ is  very small we can expand the numerator $(1+ \kappa \eta^2 p^2)^{-\frac{1}{2}}$ to first order in $\kappa\eta^2$, to obtain,
\begin{equation*}
\begin{split}
    D(t, r) = &\zeta \int\limits_0^\infty \dd p ~p^{\frac{n-3}{2}} J_{\frac{n-3}{2}}(p r)e^{-ip(1+\eta^2p^2)^{1/2} t} \\ - &\zeta\frac{\kappa\eta^2}{2} \int\limits_0^\infty \dd p ~p^{\frac{n+1}{2}}J_{\frac{n-3}{2}}(p r)e^{-ip(1+\eta^2p^2)^{1/2} t}
\end{split}
\end{equation*}
One can now expand the $(1+\kappa\eta^2 p^2)^{\frac{1}{2}}$ in the exponential, 
\begin{equation*}
    e^{-ip(1+\kappa\eta^2p^2)^{1/2} t}=e^{-ipt}e^{-i\frac{\kappa\eta^2}{2} p^3 t} \text{.}
\end{equation*}
For the second factor for small $\kappa\eta^2$ we can further expand the exponential to first order, obtaining $e^{-i\frac{\kappa\eta^2}{2} p^3 t}=1-\frac{\kappa\eta^2}{2}it$.
This would give a term proportional to $t$, which we are free to disregard as it is not an invariant quantity and we are free to rotate to a frame where $r=\theta(-\sigma^2)\sqrt{r^2-t^2+i\epsilon}, t=0$ as in the $1+1$ propagator case.
This will compute the retarded or time-like massless propagator $D^-(t,r)$.

We are left with the following integrals to evaluate,
\begin{align*}
     D^-(t, r) = &\zeta \int\limits_0^\infty \dd p ~p^{\frac{n-3}{2}} J_{\frac{n-3}{2}}(p r)e^{-ipt} \\ 
     &-\zeta\frac{\kappa\eta^2}{2} \int\limits_0^\infty \dd p ~p^{\frac{n+1}{2}}J_{\frac{n-3}{2}}(p r)e^{-ipt}
\end{align*}
To solve these we analytically continue the standard integral from \cite{gradshteyn2014table} (pg. 694. 6.623 (1)),
\begin{equation}\label{eqn:stdint_bxexp}
    \int\limits_0^\infty e^{-\gamma x} J_\nu (\beta x)x^\nu ~ \dd x= \frac{(2\beta)^\nu \Gamma(\nu+\frac{1}{2})}{\sqrt{\pi}(\gamma^2+\beta^2)^{\nu+\frac{1}{2}}} \text{.}
\end{equation}
It will be convenient to subtract a small imaginary component $i\epsilon$ to $t$ to allow convergence of the integral later, replacing $t^2$ with $t^2+i \epsilon$, after discarding the term in $\BigO{\epsilon^2}$.

For the first integral we have,
\begin{equation}\label{eqn:3p1stand}
    \zeta \int\limits_0^\infty \dd p ~p^{\frac{n-3}{2}} J_{\frac{n-3}{2}}(p r)e^{-ipt} = \frac{\Gamma(\frac{n}{2}-1)}{4(\pi)^{\frac{n}{2}}(r^2 - t^2 +i \epsilon)^{\frac{n}{2}-1}} \text{.}
\end{equation}

The second integral is more problematic, but can be massaged into standard forms by taking advantage of the differential recurrence relationship of the Bessel functions and integrating by parts.
We note,
\begin{equation}\label{eqn:bessel_recursion}
    \dv{}{x} \left \{ x^\eta J_\eta(\beta x) \right \}= \beta x^\eta J_{\eta-1}(\beta x ) \text{.}
\end{equation}

One can now write the integrand as,
\begin{equation*}
\begin{split}
    &\zeta\frac{\kappa\eta^2}{2} \int\limits_0^\infty \dd p ~p^{\frac{n+1}{2}}J_{\frac{n-3}{2}}(p r)e^{-ipt} \\
    &=\zeta\frac{\kappa\eta^2}{2r} \int\limits_0^\infty \dd p ~p e^{-ipt} \dv{}{p}\left \{ p^{\frac{n-1}{2}}J_{\frac{n-1}{2}}(p r)\right \} \text{.}
\end{split}
\end{equation*}

This is in the form of $\int u \dd v = uv-\int v \dd u$, with $u=p e^{-ipt}$ and $v=p^{\frac{n-1}{2}}J_{\frac{n-1}{2}}(p r)$.
Rotation of the time axis by an arbitrary small angle $\delta \leq \pi/2$, such that $t->\infty e^{-i\delta}$ in the upper limit of the integral, ensures the convergence of $[ uv ]^\infty_0$, and we are left with,
\begin{align*}
    & \zeta\frac{\kappa\eta^2}{2r} \int\limits_0^\infty \dd p ~p e^{-ipt} \dv{}{p}\left \{ p^{\frac{n-1}{2}}J_{\frac{n-1}{2}}(p r)\right \} = \\
    &-\zeta\frac{\kappa\eta^2}{2r} \int\limits_0^\infty \dd p~ e^{-ipt} p^{\frac{n-1}{2}}J_{\frac{n-1}{2}}(p r) \\
    &+\zeta\frac{\kappa\eta^2}{2r} i t \int\limits_0^\infty \dd p ~e^{-ipt} p^{\frac{n+1}{2}}J_{\frac{n-1}{2}}(p r) \text{.}
\end{align*}
The second of these integrals can be discarded due to the leading factor of $it$ as above, and the first is in the standard form used earlier.
Using the formula \Cref{eqn:stdint_bxexp} we have the result for this correction to order $\eta^2$,
\begin{equation}\label{eqn:3p1corr}
    \zeta\frac{\kappa\eta^2}{2} \int\limits_0^\infty \dd p ~p^{\frac{n+1}{2}}J_{\frac{n-3}{2}}(p r)e^{-ipt} = \frac{\kappa\eta^2 \Gamma(\frac{n}{2})}{4(\pi)^{\frac{n}{2}}(r^2-t^2 + i\epsilon)^{\frac{n}{2}} } \text{.}
\end{equation}

Substituting in $n=4$ into \Cref{eqn:3p1stand,eqn:3p1corr}, with $\sigma^2=t^2-r^2$ being the invariant interval, we obtain our final result for the propagator to first order in $\kappa$,
\begin{equation}\label{eqn:final31prop_spacelike}
    D^-(t,r)=\frac{\theta(-\sigma^2)}{4\pi^2(-\sigma^2 + i\epsilon)}\left \{ 1+\frac{\kappa\eta^2}{(-\sigma^2+i\epsilon)}\right \} \text{.}
\end{equation}
This has the corresponding advanced or time-like counterpart,
\begin{equation}\label{eqn:final31prop_timelike}
    D^+(t,r)=\frac{-\theta(\sigma^2)}{4\pi^2(\sigma^2 - i\epsilon)}\left \{ 1-\frac{\kappa\eta^2}{(\sigma^2-i\epsilon)}\right \} \text{.}
\end{equation}

\subsection{Second method of computation}
\label{sec:app_b2}
We start with \Cref{eqn:base_integral}, however we proceed differently.
Instead we make the simplification $p(1+\kappa\eta^2p^2)^{1/2} \sim \sqrt{\kappa}\eta p^2$ in the exponential, we can also expand the $(1+\kappa\eta^2p^2)$ for small $\kappa\eta^2$ in the denominator to $(1-\frac{\kappa\eta^2}{2}p^2 + \dots + \mathcal{O}(\kappa^2\eta^4))$, and following a change of variable $x=p^2$, we have,
\begin{equation*}
     D(t, r) = \frac{\zeta}{2}\int\limits_0^\infty \dd x ~\left (x^{\frac{n-5}{4}} -\frac{\kappa\eta^2}{2}x^{\frac{n-1}{4}} \right )J_{\frac{n-3}{2}}( r\sqrt{x})e^{-i\sqrt{\kappa}\eta x t} \text{.}
\end{equation*}

These gives two standard integrals \cite{gradshteyn2014table} (pg. 701. 6.643(i)) of the form,
\begin{equation}
\begin{split}
    \int\limits_0^\infty \dd x  ~ x^{\mu-1/2} e^{-a x} J_{2\nu}(2b\sqrt{x}) &= \\
    \frac{\Gamma(\mu + \nu +1/2)} {b\Gamma(2\nu +1)} \exp \frac{-b^2}{2 a} a^{-\mu}M_{\mu,v}\left (\frac{b^2}{a} \right ) \text{,}
\end{split}
\end{equation}
where $M_{\mu,v}(x)$ is the Whittaker special function, and we  analytically continue to imaginary exponents.
For our integrals,
\begin{align*}
    a&=i \sqrt{\kappa}\eta t \text{,}\\
    b&=\frac{r}{2} \text{,}
\end{align*}
and we have and two pairs of parameters $\mu,\nu$.
These are
\begin{align*}
    \mu&=\frac{n-3}{4} \text{,} \nu=\frac{n-3}{4} \text{, for}~ x^{\frac{n-5}{4}} \text{, and,} \\
    \mu&=\frac{n+1}{4} \text{,} \nu=\frac{n-3}{4} \text{, for} ~x^{\frac{n-1}{4}} \text{.}
\end{align*}
Completing the substitutions we have for our Green's function,
\begin{widetext}
\begin{equation}
\begin{split}
  D(t, r) &= \frac{\zeta}{r} \exp{\frac{-r^2}{8i \sqrt{\kappa}\eta t}} \times \\
  &\left [\frac{\Gamma(\frac{n-2}{2})}{\Gamma(\frac{n-1}{2})}(i \sqrt{\kappa}\eta t)^{-(\frac{n-3}{4})} M_{\frac{n-3}{4},\frac{n-3}{4}}\left ( \frac{r^2}{4i \sqrt{\kappa}\eta t}\right ) - \frac{\kappa\eta^2\Gamma(\frac{n}{2})}{2\Gamma(\frac{n-1}{2})}(i \sqrt{\kappa}\eta t)^{-(\frac{n+1}{4})} M_{\frac{n+1}{4},\frac{n-3}{4}}\left ( \frac{r^2}{4i \sqrt{\kappa}\eta t}\right ) \right ]
\end{split}
\end{equation}
\end{widetext}
This unpleasant looking result collapses rather nicely in the limit of $t \rightarrow 0$, when the argument of the Whittaker function becomes infinite. 
The Whittaker function $M_{\mu,\nu}(z)$ has the following asymptotic behavior in the case of $z \rightarrow \infty$,
\begin{equation}
   \lim_{z->\infty} M_{\mu,\nu}(z) \sim \frac{\Gamma(1+2\nu)}{\Gamma(\frac{1}{2}+ \nu-\mu)}e^{\frac{1}{2}z}z^{-\mu}\text{.}
\end{equation}
The  exponential term is the reciprocal of the $\exp{\frac{-r^2}{8i\eta t}}$ term before the brackets, and the $z^{-\mu}$ cancels the $(i\sqrt{\kappa}\eta t)^\mu$ terms preceding the Whittaker functions, leaving the $\frac{r^2}{4}$ factors.
In this limit therefore, the Green's function becomes,
\begin{equation}
\begin{split}
    &D(t, r) = \\
    &\frac{\zeta}{r} \left [ \frac{\Gamma(\frac{n-2}{2})}{\Gamma(\frac{1}{2})} \left ( \frac{r^2}{4}\right )^{-(\frac{n-3}{4})} - \frac{\kappa\eta^2}{2}\frac{\Gamma(\frac{n}{2})}{\Gamma(-\frac{1}{2})} \left ( \frac{r^2}{4}\right )^{-(\frac{n+1}{4})} \right ] \text{.}
\end{split}
\end{equation}
When taking the limit $t \rightarrow 0$, we should interpret $r^2$ as being the spacelike interval, $r^2 = \theta(-\sigma^2)(-\sigma^2 + i \epsilon)$, and the propagator as the retarded or space-like propagator $D^-(t,r)$.
Setting $n=4$ and making this substitution for $r$ we have,
\begin{equation}
     D^-(t, r) = \frac{\theta(-\sigma^2)}{4\pi^2(-\sigma^2+i\epsilon)}\left [1+ \frac{\kappa\eta^2}{(-\sigma^2+i\epsilon)} \right ] \text{.}
\end{equation}
We can further make the substitution $(-\sigma^2 + i\epsilon) \rightarrow -(\sigma^2 - i \epsilon)$ to recover the advanced or time-like propagator,
\begin{equation}
     D^+(t, r) = \frac{-\theta(\sigma^2)}{4\pi^2(\sigma^2-i\epsilon)}\left [1- \frac{\kappa\eta^2}{(\sigma^2-i\epsilon)} \right ] \text{.}
\end{equation}
It will be noted that this result is identical to that obtained using the first method in \Cref{sec:app_b1}.

\section{Detector function calculation}
\label{sec:appendix_c}
To compute the transition probability we need to perform the integral in \Cref{eqn:detector_integral}, and follow a calculation similar to that in \cite{birrell1984quantum}.
To make progress we use the modified $3+1$ propagator \Cref{eqn:final31spacelike}, choosing the advanced propagator (space-like) form,
\begin{equation*}
\begin{split}
    D^{+}(\Delta t,\Delta z)=&-\frac{1}{4\pi^2(\Delta t^2-\Delta z^2 - i\epsilon)} \times \\
    &\left \{ 1- \frac{\kappa \eta^2}{(\Delta t^2-\Delta z^2-i\epsilon)}\right \} \text{.}
\end{split}
\end{equation*}

Using the elementary properties of the hyperbolic functions, we have for $\Delta t^2$ and $\Delta z^2$,
\begin{align*}
    \Delta t &= 2\alpha \cosh{\left( \frac{\tau+\tau'}{2\alpha}\right )} \sinh{\left( \frac{\tau-\tau'}{2\alpha}\right )}\\
    \Delta z &= 2\alpha \sinh{\left( \frac{\tau+\tau'}{2\alpha}\right )} \sinh{\left( \frac{\tau-\tau'}{2\alpha}\right )}
\end{align*}
Inserting these back into the propagator we have,
\begin{equation}\label{eqn:hyperbolic_prop}
\begin{split}
    D^{+}(\Delta \tau)=&- \frac{1}{16 \pi^2\alpha^2} \sinh^{-2} \left ( \frac{\Delta \tau}{2\alpha } - \frac{i\epsilon}{\alpha}\right )\times\\
    &\left [ 1- \frac{\kappa \eta^2}{4\alpha^2}\sinh^{-2} \left ( \frac{\Delta \tau}{2\alpha } - \frac{i\epsilon}{\alpha}\right ) \right ]  \text{.}
\end{split}
\end{equation}
One can make use of the identities,
\begin{align*}
    \sum\limits_{k=-\infty}^{\infty} (x-\pi i k )^{-2} &= \sinh^{-2} x \text{,}\\
    \sum\limits_{k=-\infty}^{\infty} (x-\pi i k )^{-4} &= \frac{1}{3}( \tanh^{-2} x \sinh^{-2} x + \sinh^{-4} x) \text{,}
\end{align*}
to allow us to restate the $\sinh^{-4}$ term as,
\begin{equation}\label{eqn:csch4}
    \sinh^{-4} x = \sum\limits_{k=-\infty}^{\infty} (x-\pi i k )^{-4} - \frac{2}{3} \sum\limits_{k=-\infty}^{\infty} (x-\pi i k )^{-2}  \text{,}
\end{equation}
and then write the propagator as,
\vspace{0.25cm}
\begin{widetext}
\begin{equation}\label{eqn:propsum}
    D^{+}(\Delta \tau)=-\frac{1}{4\pi^2}\left (1+\frac{\kappa \eta^2}{6\alpha^2}\right)\sum\limits_{k=-\infty}^\infty (\Delta \tau -2i \epsilon -2\pi i \alpha k)^{-2} + \frac{\kappa \eta^2}{4\pi^2}\sum\limits_{k=-\infty}^\infty (\Delta \tau -2i \epsilon -2\pi i \alpha k)^{-4}
\end{equation}
\end{widetext}
This expression for the propagator can be substituted back into \Cref{eqn:transprob}, and evaluated by reversing the order of the sum and the integral.
To evaluate  the integral we choose a semi-circular contour around the multiple pole at $\Delta \tau = 2\pi i \alpha k$, closed in the upper half plane.
The integral then evaluates to $2\pi i$ times the residue at $2\pi i \alpha k$.
These residues are found by elementary means to be $-i(E-E_0)e^{2 \pi \alpha(E-E_0)k}$, and $\frac{i}{6}(E-E_0)e^{2 \pi \alpha(E-E_0)k}$ respectively.
We have for our two integrals the values,
\begin{widetext}
\begin{align*}
    &\int\limits_{-\infty}^\infty \dd (\Delta \tau) \frac{e^{-i(E-E_0)\Delta \tau}}{(\Delta \tau -2i \epsilon -2\pi i \alpha k)^{2}} = 2\pi(E-E_0)e^{2\pi(E-E_0)\alpha k}\\
    &\int\limits_{-\infty}^\infty \dd (\Delta \tau) \frac{e^{-i(E-E_0)\Delta \tau}}{(\Delta \tau -2i \epsilon -2\pi i \alpha k)^{4}} = -\frac{\pi}{3}(E-E_0)^3e^{2\pi(E-E_0)\alpha k}
\end{align*}
\end{widetext}
The sum over $k$ is from $k=-\infty$ to $k=\infty$, but our choice of contour for the integral subtracts the contribution to the sum from $k \in [-\infty,0)$ leaving the result as the sum of a geometric series.
One obtains for the transition probability,
\begin{widetext}
\begin{equation}
    \frac{c^2}{2 \pi}\sum \limits_E \abs{\bra{E}m(0) \ket{E_0}}^2 \left \{ \left ( 1+\frac{\kappa\eta^2}{6 \alpha^2}\right ) \frac{E-E_0}{e^{2\pi(E-E_0)\alpha} -1 } + \left (\frac{\kappa\eta^2}{6}\right) \frac{(E-E_0)^3}{e^{2\pi(E-E_0)\alpha} -1 } \right \} \text{.}
\end{equation}
And correspondingly for the detector response function,
\begin{equation}\label{eqn:detector_final_appendix}
    \mathcal{F}(E) = \left \{ \left ( 1+\frac{\kappa\eta^2}{6 \alpha^2}\right ) \frac{E-E_0}{e^{2\pi(E-E_0)\alpha} -1 } + \left (\frac{\kappa\eta^2}{6}\right) \frac{(E-E_0)^3}{e^{2\pi(E-E_0)\alpha} -1 } \right \} \text{.}
\end{equation}
\end{widetext}

The Planck factor is unaffected by the modification to the propagator, and the presence of this factor suggests that the particles detected by the Unruh detector are essentially thermal.
This can be further reinforced by examining the periodic behavior of the Green's function.
It is well known \cite{birrell1984quantum} that a thermal Green's function $G_\beta$ is periodic in imaginary time.
In particular it satisfies $G_\beta^{\pm}(t,\vb{x};t',\vb{x}') = G_\beta^{\mp}(t+i\beta,\vb{x};t',\vb{x}')$, where $\beta = (kT)^{-1}$.
It will be noted that for our smoothly accelerating detector the propagator \Cref{eqn:hyperbolic_prop} is composed of terms in $\sinh^{-2}(\Delta t)$.
It is well known that the hyperbolic functions are periodic with an imaginary period of $2 \pi i$, however, both of the $\sinh$ terms are squared.
This has the effect of changing the period to $\pi i$.
Inspection of \Cref{eqn:hyperbolic_prop}, reveals that the period of the propagator is $2\pi\alpha$, indicating that $\beta=2\pi\alpha$ as suggested by the Planck factor in \Cref{eqn:detector_final}.

\section{Computation of Ho{\u r}ava-Lifshitz Corrections}
\label{sec:appendix_d}

As mentioned in \Cref{sec:introduction}, and discussed in \Cref{sec:Horava_lifshitz}, the Ho{\u r}ava-Lifshitz theory of gravity can introduce higher order terms in $p$ into the dispersion relationship.
In general the $p^6$ term has opposite sign to the $p^4$ term, and so to asses the effect of light speed propagation on the result obtained in \Cref{sec:accelerated}, we propose for a massless particle the following dispersion relation,
\begin{equation}
    E^2=p^2+\kappa \eta^2 p^4 - \kappa \eta^4 p^6 \text{,}
\end{equation}
where $\eta^4$ is introduced on dimensional grounds.

We repeat our calculation in $3+1$ dimensions to obtain the $D^+(t,r)$ propagator as outlined in \Cref{sec:app_b1}.
The crux of the calculation involves expanding the denominator of the integral measure, $E_p= \sqrt{p^2+\kappa \eta^2 p^4 - \kappa \eta^4 p^6}$, for small $\kappa \eta^2$.
We proceed as before, but  collect terms up to $p^4$, and as such we now have,
\begin{equation*}
    (E_p)^{-1} \simeq p \left [1-\frac{1}{2}\kappa \eta^2 p^2 + \frac{1}{2}\kappa\left ( \frac{3}{4}\kappa + 1 \right ) \eta^4 p^4 + \dots \right ] \text{.}
\end{equation*}
For brevity we will refer to $\kappa'=\kappa\left ( \frac{3}{4}\kappa + 1 \right )$, but note for $\kappa=1, \kappa'=\frac{7}{4}$, and $\kappa=-1, \kappa'=-\frac{1}{4}$, preserving the opposite polarity of the effect to the $p^4$ term.

Following the analysis in \Cref{sec:app_b1}, the expression for the space-like propagator acquires an additional integral contribution,
\begin{align*}
     D^-(t, r) = ~&\zeta \int\limits_0^\infty \dd p ~p^{\frac{n-3}{2}} J_{\frac{n-3}{2}}(p r)e^{-ipt} \\ 
     &-\zeta\frac{\kappa\eta^2}{2} \int\limits_0^\infty \dd p ~p^{\frac{n+1}{2}}J_{\frac{n-3}{2}}(p r)e^{-ipt} \\
     &+\zeta\frac{\kappa'\eta^4}{2} \int\limits_0^\infty \dd p ~p^{\frac{n+5}{2}}J_{\frac{n-3}{2}}(p r)e^{-ipt} \text{,}
\end{align*}
after discarding the irrelevant contribution from expanding $E_p$ in the exponential.

We can proceed identically for the third integral as undertaken for the second, making use of the Bessel function recursion relationship \Cref{eqn:bessel_recursion} twice.
After manipulation the third integral is reduced to,
\begin{equation*}
    \frac{3\zeta\kappa' \eta ^4}{2r^2} \int\limits_0^\infty \dd p ~p^{\frac{n+1}{2}}J_{\frac{n+1}{2}}(p r)e^{-ipt} \text{,}
\end{equation*}
whose value can be read off from the standard integral \Cref{eqn:stdint_bxexp}, as,
\begin{equation*}
    \frac{3\Gamma(\frac{n+2}{2})\kappa' \eta^4}{2(\pi)^{\frac{n}{2}}(r^2-t^2+i\epsilon)^{\frac{n+2}{2}}} \text{.}
\end{equation*}
Setting $n=4$, this can now be inserted into the expression for the propagator, and for the time-like advanced propagator we have,
\begin{equation}\label{eqn:final31prop_Horava}
    D^+(t,r)=\frac{-\theta(\sigma^2)}{4\pi^2(\sigma^2 - i\epsilon)}\left \{ 1-\frac{\kappa\eta^2}{(\sigma^2-i\epsilon)} + \frac{12 \kappa' \eta^4}{(\sigma^2 - i \epsilon)^2}\right \} \text{.}
\end{equation}
We note the positive sign for the $(\sigma^2 - i \epsilon)^{-2}$ term, which arises from the conversion of $(-\sigma^2 + i\epsilon)$ to $(\sigma^2 - i\epsilon)$ inside an overall cubic denominator. 

Turning to the computation of the detector response function, we follow the calculation in \Cref{sec:appendix_c}, noting that upon inserting the smoothly accelerating trajectory we had an additional term in the propagator,
\begin{equation}\label{eqn:hyperbolic_Horava_prop}
\begin{split}
    &D^{+}(\Delta \tau)=- \frac{1}{16 \pi^2\alpha^2} \sinh^{-2} \left ( \frac{\Delta \tau}{2\alpha } - \frac{i\epsilon}{\alpha}\right )\times\\
    &\left [ 1- \frac{\kappa \eta^2}{4\alpha^2}\sinh^{-2} \left ( \frac{\Delta \tau}{2\alpha } - \frac{i\epsilon}{\alpha}\right )  + \frac{3\kappa' \eta^4}{4\alpha^4}\sinh^{-4} \left ( \frac{\Delta \tau}{2\alpha } - \frac{i\epsilon}{\alpha}\right ) \right ]  \text{.}
\end{split}
\end{equation}
Similar considerations to \Cref{eqn:csch4} allow us to substitute for the $\sinh^{-6}$ term the following infinite sum,
\begin{widetext}
\begin{equation}\label{eqn:csch6}
    \sinh^{-6} x =  \sum\limits_{k=-\infty}^{\infty} (x-\pi i k )^{-6} - \sum\limits_{k=-\infty}^{\infty} (x-\pi i k )^{-4} - \frac{4}{5}\sum\limits_{k=-\infty}^{\infty} (x-\pi i k )^{-2}  \text{.}
\end{equation}
The calculation proceeds identically and yields a similar result with additional correction terms at all orders of $(E-E_0)$ in the detector response function,

\begin{equation}\label{eqn:detector_Horava_appendix}
    \mathcal{F}(E) = \left \{ \left ( 1+\frac{\kappa\eta^2}{6 \alpha^2} -\frac{3\kappa'\eta^4}{5\alpha^4}\right ) \frac{E-E_0}{e^{2\pi(E-E_0)\alpha} -1 } + \left (\frac{\eta^2[\kappa+12\kappa'\eta^2]}{6}\right) \frac{(E-E_0)^3}{e^{2\pi(E-E_0)\alpha} -1 } + \left (\frac{\kappa'\eta^4}{10}\right) \frac{(E-E_0)^5}{e^{2\pi(E-E_0)\alpha} -1 } \right \} \text{.}
\end{equation}
\end{widetext}
It was remarked in \Cref{sec:appendix_c} that the propagator, being stated in terms of $\sinh^{-2}$, is periodic with a period of $\pi i$, which strengthens the interpretation of $2\pi\alpha$ as the temperature of the radiation.
Although we now have a term in $\sinh^{-4}$, the period is the same and we can conclude that the propagator \Cref{eqn:hyperbolic_Horava_prop} satisfies $G_\beta^{\pm}(t,\vb{x};t',\vb{x}') = G_\beta^{\mp}(t+i\beta,\vb{x};t',\vb{x}')$, where $\beta = (kT)^{-1}=2\pi\alpha$.
%\nocite{*}

\bibliography{PixelatedDetectors}% Produces the bibliography via BibTeX.

\end{document}